\documentclass[aps,showpacs,preprintnumbers,amsmath,amssymb]{revtex4}
\usepackage{dcolumn}
\usepackage[dvips]{epsfig}

\usepackage{float}
\usepackage{graphics,epsfig}
\usepackage{graphicx}
\usepackage{dcolumn}
\usepackage{bm}
\usepackage{gensymb}

\begin{document}
\baselineskip=0.8 cm
\title{\bf Polarization effects in Kerr black hole shadow due to the coupling between  photon and bumblebee field}

\author{Songbai Chen$^{1,3}$\footnote{Corresponding author: csb3752@hunnu.edu.cn},
Mingzhi Wang$^{2}$\footnote{wmz9085@126.com}
Jiliang Jing$^{1,3}$ \footnote{jljing@hunnu.edu.cn}}
\affiliation{ $ ^1$ Department of Physics, Key Laboratory of Low Dimensional Quantum Structures
and Quantum Control of Ministry of Education, Synergetic Innovation Center for Quantum Effects and Applications, Hunan
Normal University,  Changsha, Hunan 410081, People's Republic of China
\\
$ ^2$ School of Mathematics and Physics, Qingdao University of Science and Technology,
Qingdao, Shandong 266061, People¡¯s Republic of China
\\
$ ^3$Center for Gravitation and Cosmology, College of Physical Science and Technology, Yangzhou University, Yangzhou 225009, People's Republic of China}

\begin{abstract}
\baselineskip=0.6 cm
\begin{center}
{\bf Abstract}
\end{center}
We present firstly the equation of motion for the photon coupled to a special bumblebee vector field in a Kerr black hole spacetime and find that the propagation of light depends on its polarization due to the birefringence phenomenon. The dependence of black hole shadow on the light's polarization is dominated by the rotation of black hole. In the non-rotating case, we find that the black hole shadow is independent of the polarization of light.  However, the status is changed in the rotating case,  in which the black hole shadow depends on the light's polarization and the coupling between bumblebee vector field and electromagnetic field. These features of black hole shadow casted by polarized lights could help us to understand the bumblebee vector field with Lorentz symmetry breaking and its interaction with electromagnetic field.

\end{abstract}

 \maketitle
\newpage
\section{Introduction}

One of the most exciting events in observing black holes is that Event Horizon Telescope Collaboration \cite{fbhs1,fbhs6} released the first image of the supermassive black hole in the center of M87 galaxy at the last year. The information stored in the image has been applied extensively to study the possibility of constraining black hole parameters and extra dimension size \cite{extr1,extr2}, and to probe some fundamental physics issues including dark matter \cite{tomoch,dark1,dark2,dark3,dark4}, the equivalence principle \cite{epb}, and so on.  Black hole shadow, caused by light rays that fall into an event horizon,  is one of  the most important ingredients in the image. Generally, the black hole shadow in the observer's sky is determined by black hole, the propagation of light ray and the position of observer. It is shown that black hole shadow depends also on the polarization of light itself if we consider some interactions between electromagnetic and gravitational fields, such as Weyl tensor coupling \cite{sb1}, which can be attributed to the birefringence phenomenon of light originating from such kinds of couplings. This interesting feature could trigger the further study of black hole shadows  under interactions between electromagnetic and other fields.

It is well known that Lorentz invariance has been great of importance in many fields of the fundamental physics. However, the development of unified gauge theories and the signals from high energy cosmic rays \cite{lvia1,lvia2} imply that the spontaneous breaking of Lorentz symmetry may emerge in the more fundamental physics defined in a higher scale of energy. Generally, it is currently impossible directly to
test such kind of physical theories with Lorentz violation  through experimentation due to their higher energy scale. However, it is possible that some signals associated with the breaking of Lorentz symmetry can emerge at sufficiently low energy scales and their corresponding effects could be observed in experiments at current energy scales \cite{casa}. Bumblebee model \cite{kost} is a simple effective theory of gravity with Lorentz violation in the standard model extension. In this theoretical model, the spontaneous breaking of Lorentz symmetry is induced by a nonzero vacuum expectation value of a bumblebee field $B_{\mu}$ under a suitable potential. Recently, the bumblebee gravity model has been extensively studied in literature \cite{ber,kost2,blum, kost3,seif,malu,guio,esco,assu}.  An exact Schwarzschild-like
black hole solution in this bumblebee gravity model is obtained  and some classical tests are studied \cite{casa}. The gravitational deflection angle of light \cite{ovgu} and the Hawking radiation \cite{kanz} are studied in this black hole spacetime. Moreover, this black hole solution is generalized to the rotation case and the corresponding  shadow \cite{ding}, accretion disk \cite{ding1} and the deflection of particle \cite{liz} around the black hole are studied. The traversable wormhole solution is also found in the framework of the bumblebee gravity theory \cite{ovgu2} and the cosmological
implications of bumblebee gravity model are further investigated in \cite{cape}. These analysis may be useful in testing the bumblebee gravity model and detecting the effects originating from the spontaneous Lorentz symmetry breaking.

From the previous analysis, the interaction between electromagnetic field and gravitational field will modify Maxwell equation and lead to that the propagation of photon in the curved spacetime depends on its polarization direction. The recent investigation shows that the coupling between axion and photon also results in the photon birefringence as photon crosses over axion matter. Axion is a compelling dark matter candidate and provides an elegant solution to the strong $CP$ problem. The axion dark matter distribution near M87 black hole \cite{tomoch} and in the protoplanetary disk around a young star \cite{tomo} are analyzed by using the birefringence effects of the polarization photons. The polarization-dependent bending that a ray of light
experiences by traveling through an axion cloud is studied in the background of a Kerr black hole \cite{Alexis}.
However, it is an open issue how the polarized light affect the shadow of a rotating black hole. In this paper, we will consider the coupling between photon and bumblebee field to study what effect of the corresponding birefringence on the shadow of a Kerr black hole. The main reason why we here chose bumblebee field is that the equation of motions for the photons with different polarizations are not coupled together in this model, which is very  critical to analyze the propagation paths of the different polarization photons and to probe the properties of black hole shadow. Moreover, we also want to see whether black hole shadow contains the information on Lorentz symmetry breaking.

The plan of our paper is organized as follows: In Sec.II,  we derive
equations of motion for the photons coupled to bumblebee vector field in a Kerr black hole spacetime from the modified Maxwell equation in the geometric optics approximation
\cite{Drummond,Daniels,Daniels1, Caip,Cho1,Lorenci,Lorenci1,Lorenci2}. In Sec.III, we study how Kerr black hole shadow change with the polarization of light and the coupling parameter under the photon-bumblebee interaction. We end the paper with a summary.

\section{Equation of motion for the photons coupled to bumblebee  field in a Kerr black hole spacetime}

In this section, we will make use of the geometric optics approximation
\cite{Drummond,Daniels,Daniels1,Caip,Cho1,Lorenci,Lorenci1,Lorenci2} and get
the equations of motion for the photons interacting with bumblebee vector field
in a Kerr black hole spacetime.
The simplest action of the electromagnetic field coupled to bumblebee vector field in the curved spacetime can be expressed as
\begin{eqnarray}
S=\int d^4x \sqrt{-g}\bigg[\frac{R}{16\pi
G}-\frac{1}{4}\bigg(F_{\mu\nu}F^{\mu\nu}+B_{\mu\nu}B^{\mu\nu}-4\alpha
B^{\mu}B^{\rho}F_{\mu\nu}F_{\rho}^{\;\nu}\bigg)-V(B^{\mu})\bigg],\label{acts}
\end{eqnarray}
where $B_{\mu}$ is bumblebee field and the corresponding strength tensor $B_{\mu\nu}$  is defined by \cite{kost}
\begin{eqnarray}
B_{\mu\nu}=\partial_{\mu}B_{\nu}-\partial_{\nu}B_{\mu}.
\end{eqnarray}
The potential $V(B^{\mu})$, inducing Lorentz violation, can be expressed as \cite{kost,casa, ber,kost2,blum, kost3,seif,malu,guio,esco,assu}
\begin{eqnarray}
V=V(B_{\mu}B^{\mu}\pm b^2),
\end{eqnarray}
where $b^2$ is a positive constant. In order to ensure the breaking of the $U(1)$ symmetry, the potential $V(B_{\mu}B^{\mu}\pm b^2)$ is supposed to own a minimum at $B_{\mu}B^{\mu}=\mp b^2$. The condition $B_{\mu}B^{\mu}= \mp b^2$ is satisfied when the vector field has a nonzero vacuum value \cite{kost,casa, ber,kost2,blum, kost3,seif,malu,guio,esco,assu}
\begin{eqnarray}
B_{\mu}=<B_{\mu}>=b_{\mu},
\end{eqnarray}
with $b_{\mu}b^{\mu}=\mp b^2$.  $F_{\mu\nu}$ is
the usual electromagnetic tensor and $\alpha$ is a dimensionless coupling constant. With the variational method,  one can find that Maxwell equation is modified as
\begin{eqnarray}
\nabla_{\mu}\bigg(F^{\mu\nu}-2\alpha b^{\mu}b^{\rho}F_{\rho}^{\;\nu}+2\alpha b^{\nu}b^{\rho}F_{\rho}^{\;\mu}\bigg)=0,\label{WE}
\end{eqnarray}
which means that the coupling changes propagation of electromagnetic field in background spacetime. Resorting to the
geometric optics approximation, one can obtain equation of motion for a coupled photon from the corrected Maxwell equation (\ref{WE}). In the geometric optics approximation,  the wavelength of photon $\lambda$ is assumed to be much smaller than a typical curvature scale $L$, but
larger than the electron Compton wavelength $\lambda_e$. This ensures that we can neglect the change of the
background gravitational and electromagnetic fields with the typical
curvature scale for the photon propagation
\cite{Drummond,Daniels,Daniels1,Caip,Cho1,Lorenci,Lorenci1,Lorenci2}. With this approximation, the electromagnetic field strength can be written as a simpler form
\begin{eqnarray}
F_{\mu\nu}=f_{\mu\nu}e^{i\theta},\label{ef1}
\end{eqnarray}
where $f_{\mu\nu}$ is a slowly varying amplitude and $\theta$ is a rapidly varying phase. It implies that the derivative term $f_{\mu\nu;\lambda}$ is not dominated so that it can be neglected in this case. The wave vector is $k_{\mu}=\partial_{\mu}\theta$, which can be treated as the usual photon momentum in quantum theories. With the help of the Bianchi identity
\begin{eqnarray}
D_{\lambda} F_{\mu\nu}+D_{\mu} F_{\nu\lambda}+D_{\nu} F_{\lambda\mu}=0,
\end{eqnarray}
one can obtain  the constraint for the amplitude $f_{\mu\nu}$, i.e.,
\begin{eqnarray}
k_{\lambda}f_{\mu\nu}+k_{\mu} f_{\nu\lambda}+k_{\nu} f_{\lambda\mu}=0.
\end{eqnarray}
It means that $f_{\mu\nu}$ can be further expressed as
\begin{eqnarray}
f_{\mu\nu}=k_{\mu}a_{\nu}-k_{\nu}a_{\mu},\label{ef2}
\end{eqnarray}
with a polarization vector $a_{\mu}$ satisfying the condition that
$k_{\mu}a^{\mu}=0$.
Inserting Eqs.(\ref{ef1}) and (\ref{ef2}) into Eq.(\ref{WE}), we
can get that the equation of motion for the photon coupled with bumblebee vector field
\begin{eqnarray}
k_{\mu}k^{\mu}a^{\nu}-2\alpha
b^{\rho}b^{\nu}k^{\mu}k_{\mu}a_{\rho}+2\alpha
b^{\rho}b^{\nu}k_{\rho}k_{\mu}a^{\nu}+2\alpha
b^{\rho}b^{\nu}k^{\nu}k_{\mu}a_{\rho}=0.\label{WE2}
\end{eqnarray}
It is obvious that the interaction with bumblebee field  affects the
propagation of the coupled photon in the background spacetime.

The Kerr metric describes the geometry of a rotation black hole, whose line-element can be expressed as
\begin{eqnarray}
ds^2&=&-\rho^2\frac{\Delta}{\Sigma^2}dt^2+\frac{\rho^2}{\Delta}dr^2+\rho^2
d\theta^2+\frac{\Sigma^2}{\rho^2}\sin^2{\theta}(d\phi-\omega dt)^2,\label{m1}
\end{eqnarray}
with
\begin{eqnarray}
\omega&=&\frac{2aMr}{\Sigma^2},
\;\;\;\;\;\;\;\;\;\;\;\;\;\;\;\;\;\;\;\;\;\;\;\;\;\;\;\;\;\rho^2=r^2+a^2\cos^2\theta,
\nonumber\\
\Delta&=&r^2-2Mr+a^2,\;\;\;\;\;\;\;\;\;\;\;\;\;\;\;\;\Sigma^2=(r^2+a^2)^2-a^2\sin^2\theta \Delta.
\end{eqnarray}
Here the parameters $M$ and $a$ denote the mass and the angular momentum per unit mass of the black hole, respectively.
In order to introduce a local set of orthonormal frames in Kerr black hole spacetime, one can use the vierbein fields defined by
\begin{eqnarray}
g_{\mu\nu}=\eta_{ab}e^a_{\mu}e^b_{\nu},
\end{eqnarray}
where $\eta_{ab}$ is the Minkowski metric and the vierbeins
\begin{eqnarray}
e^a_{\mu}=\left(\begin{array}{cccc}
\rho\frac{\sqrt{\Delta}}{\Sigma}&0&0&-\frac{\omega\Sigma}{\rho}\sin\theta\\
0&\frac{\rho}{\sqrt{\Delta}}&0&0\\
0&0&\rho&0\\
&0&0&\frac{\Sigma}{\rho}\sin\theta
\end{array}\right),
\end{eqnarray}
with the inverse
\begin{eqnarray}
e_a^{\mu}=\left(\begin{array}{cccc}
\frac{\Sigma}{\rho\sqrt{\Delta}}&0&0&0\\
0&\frac{\sqrt{\Delta}}{\rho}&0&0\\
0&0&\frac{1}{\rho}&0\\
\frac{\omega\Sigma}{\rho\sqrt{\Delta}}&0&0&\frac{\rho}{\Sigma\sin\theta}
\end{array}\right).
\end{eqnarray}
Making use of the relationship $a^{\mu}k_{\mu}=0$, one can find that the equation of motion of the photon coupling with  (\ref{WE2}) can be simplified as a set of equations for three independent polarisation components $a^{t}$, $a^{\theta}$, and $a^{\phi}$,
\begin{eqnarray}
\bigg(\begin{array}{ccc}
K_{11}&K_{12}&K_{13}\\
K_{21}&K_{22}&
K_{23}\\
K_{31}&K_{32}&K_{33}
\end{array}\bigg)
\bigg(\begin{array}{c}
a^t\\
a^{\theta}
\\
a^{\phi}
\end{array}\bigg)=0.\label{Kk1}
\end{eqnarray}
The coefficients $K_{ij}$, ($i,j=t, \theta, \phi$) are very complicated and we do not list them here for simplicity. There exists the non-zero solution of Eq.(\ref{Kk1}) only if the determinant of the coefficient matrix $|K|$ is equal to zero (.i.e., $|K|=0$). However,  we find that in the Kerr black hole spacetime it is difficult to find a solution satisfied $|K|=0$
in a general case. Here, we focus on only a special case in which
the bumblebee field is spacelike and has the form $b^{\mu}=(0,b^{r},b^{\theta},0)$. We find that in this case there exist two solutions for equation (\ref{Kk1}), i.e.,
\begin{eqnarray}
&&-(e_{t}^0)^2k^tk^t+(e_{r}^1)^2k^rk^r+(e_{\theta}^2)^2k^{\theta}k^{\theta}
+(e_{t}^3k^t+e_{\phi}^3k^{\phi})^2
+\frac{8\alpha}{1-2\alpha b^2}(e_{r}^1)^2(e_{\theta}^2)^2b^{r}b^{\theta}k^{r}k^{\theta}=0, \label{Kcoe2}\\
&&-(e_{t}^0)^2k^tk^t+[1+2\alpha(e_{r}^1)^2b^{r}b^{r}](e_{r}^1)^2k^rk^r+
[1+2\alpha(e_{\theta}^2)^2b^{\theta}b^{\theta}](e_{\theta}^2)^2k^{\theta}k^{\theta}
+(e_{t}^3k^t+e_{\phi}^3k^{\phi})^2
\nonumber\\&&
+4\alpha(e_{r}^1)^2(e_{\theta}^2)^2b^{r}b^{\theta}k^{r}k^{\theta}=0.
\label{Kcoe3}
\end{eqnarray}
This means that the effects of the interaction with bumblebee field yields that a coupled photon propagation in a Kerr black hole spacetime depend on  the polarizations of photon, which is the so-called birefringence phenomenon.
When the coupling constant $\alpha=0$, the
interaction vanish and then the light-cone conditions (\ref{Kcoe2}) and (\ref{Kcoe3}) recover to the usual form in a Kerr spacetime in which the birefringence phenomenon vanishes because the photon propagation is independent of its polarization directions.

The light cone conditions (\ref{Kcoe2}) and (\ref{Kcoe3}) actually indicate that the motion of the coupled photons is non-geodesic in the Kerr metric. However, these photons can be looked as moving along the null geodesics of the effective metric $\gamma_{\mu\nu}$, i.e., $\gamma^{\mu\nu}k_{\mu}k_{\nu}=0$ \cite{Breton}. Obviously, the effective metric depends on the polarization of light, the bumblebee vector field and the coupling constant.
Here we focus on only the following a special case of bumblebee vector field with $b^2=1$, and
\begin{eqnarray}
B_{\mu}=(0,\;\frac{r}{\sqrt{\Delta}},\;a\cos\theta,\;0).\label{beefield}
\end{eqnarray}
It is obvious that the tensor
$B_{\mu\nu}=\partial_{\mu}B_{\nu}-\partial_{\nu}B_{\mu}=0$ for above bumblebee field and it satisfies the equation of motion
\begin{eqnarray}
\nabla_{\mu}B^{\mu\nu}=\frac{\partial V}{\partial x^{\nu}}\bigg|_{B^{\mu}B_{\mu}=1}=0.
\end{eqnarray}
We find that the effective metric for the polarized light corresponded to the equations (\ref{Kcoe2}) or (\ref{Kcoe3}) is
\begin{eqnarray}
ds^2_{I}=-\rho^2\frac{\Delta}{\Sigma^2}dt^2+
\frac{\rho^2}{\Delta W(r,\theta)}dr^2-
\frac{8\alpha ar\cos\theta}{(1-2\alpha)\sqrt{\Delta}W(r,\theta)}drd\theta
+\frac{\rho^2}{W(r,\theta)}
d\theta^2+\frac{\Sigma^2}{\rho^2}\sin^2{\theta}(d\phi-\omega dt)^2,\label{mec1}
\end{eqnarray}
or
\begin{eqnarray}
ds^2_{II}=-\rho^2\frac{\Delta}{\Sigma^2}dt^2+
\frac{\rho^2+2\alpha a^2\cos^2\theta}{(1+2\alpha)\Delta }dr^2-
\frac{4\alpha a r\cos\theta}{(1+2\alpha)\sqrt{\Delta}}drd\theta
+\frac{\rho^2+2\alpha r^2}{1+2\alpha}
d\theta^2+\frac{\Sigma^2}{\rho^2}\sin^2{\theta}(d\phi-\omega dt)^2,\label{mec2}
\end{eqnarray}
with
\begin{eqnarray}
W(r,\theta)&=&1-\frac{16\alpha^2r^2a^2\cos^2\theta}{(1-2\alpha)^2\rho^4}.
\end{eqnarray}
The cross term $\gamma_{r\theta}$ makes the effective metric more complicated than that of the usual Kerr one. Moreover, we find that the singularity occurs for the effective metrics (\ref{mec1})-(\ref{mec2}) at $\alpha=\pm\frac{1}{2}$. As $\alpha=\frac{1}{2}$, one can find the null geodesic condition in the effective metric (\ref{mec1}) becomes
\begin{eqnarray}
\gamma_{\mu\nu}\dot{x}^{\mu}\dot{x}^{\nu}=-\rho^2\frac{\Delta}{\Sigma^2}\dot{t}^2+\frac{\Sigma^2}{\rho^2}\sin^2{\theta}(\dot{\phi}
-\omega \dot{t})^2=0,
\end{eqnarray}
which implies that the dispersion relation for arbitrary photon does not depend on its four-velocity components $\dot{r}$ and $\dot{\theta}$ at the arbitrary spacetime point. Similarly, as $\alpha=-\frac{1}{2}$, the dispersion relation for arbitrary photon in the effective metric (\ref{mec2}) does not depend on its four-velocity components $\dot{t}$ and $\dot{\phi}$ at arbitrary position. This means that the propagation of polarized photon are unphysical in these two limited cases. Thus, in order to avoid these singularities  arising from the coupling in the effective metric, the coupling constant $\alpha$ must be limited in the range $\alpha\in (-\frac{1}{2},\frac{1}{2})$. For a sake of simplicity, we write the effective metrics (\ref{mec1}) and (\ref{mec2}) as a unified form
 \begin{eqnarray}
ds^2_{i}&=&-\rho^2\frac{\Delta}{\Sigma^2}dt^2+
F_i\frac{\rho^2}{\Delta}dr^2-
2H_i \frac{\rho^2}{\sqrt{\Delta}}drd\theta
+U_i\rho^2
d\theta^2+\frac{\Sigma^2}{\rho^2}\sin^2{\theta}(d\phi-\omega dt)^2,\label{meun}
\end{eqnarray}
where $F_i$,  $H_i$, and  $U_i$ are functions of coordinates $r$ and $\theta$. Obviously, there exist two conserved quantities $E$ and $L$, which corresponds to the energy and the $z$-component of the angular momentum of the polarized photon moving in the effective effective metric  (\ref{meun}). With the help of these two conserved quantities, the null geodesics for the effective metric  (\ref{meun}) can be expressed as
\begin{eqnarray}
\rho^2\dot{t}&=&\frac{(r^2+a^2)[(r^2+a^2)E-aL]}{\Delta}-a(aE\sin^2\theta-L),\label{u3}\\
\rho^2\dot{\phi}&=&\frac{a[(r^2+a^2)E-aL]}{\Delta}-(aE-\frac{L}{\sin^2\theta}),\label{u4}\\
F_i\frac{\rho^4}{\Delta}\dot{r}^2_i &-&
2H_i \frac{\rho^4}{\sqrt{\Delta}}\dot{r}_i\dot{\theta}_i
+U_i\rho^4\dot{\theta}^2_i=\frac{[(r^2+a^2)E-aL]^2}{\Delta}-
\bigg(\frac{aE\sin^2\theta-L}{\sin\theta}\bigg)^2.
\label{cedis}
\end{eqnarray}
With these equations, we can study the propagation of the polarized light in the Kerr black hole space time and probe the properties of black hole shadow casted by the polarized lights.
\begin{figure}
  \includegraphics[width=16.0cm ]{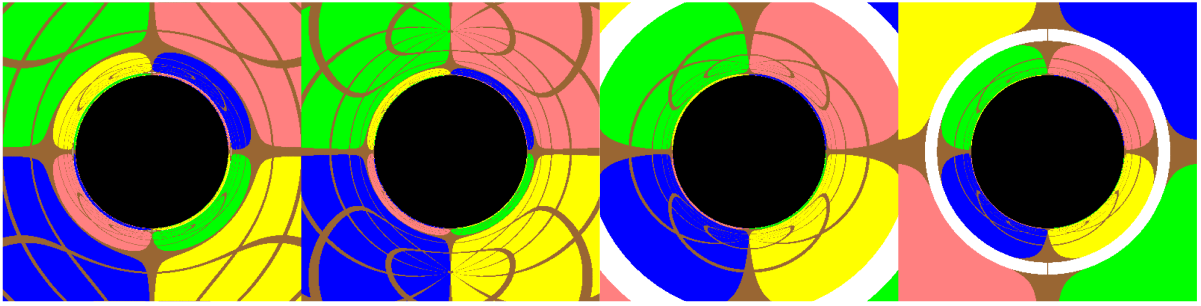}
\caption{The change of black hole shadow with the coupling constant $\alpha$ for the metric (\ref{mec2}). Here we set the black hole $M=1$ and rotation parameter $a=0$, and the observer locates in the position with $r_{obs}=50M$  and  $\theta_{obs}=\pi/2$. The figures from left to right correspond to $\alpha=-0.3$, $-0.2$, $0$, and $0.2$, respectively.}
\label{as0}
\end{figure}

\section{Kerr black hole shadow due to coupling between the photon and bumblebee vector field}

When the rotation parameter $a$ tends to zero, the metric (\ref{mec1}) reduces to that of the usual Schwarzschild black hole and is independent of the coupling parameter $\alpha$, which means that the polarized light moving along the geodesics in the effective metric (\ref{mec1}) does not affect the black hole shadow in this case. However, for the metric (\ref{mec2}), it is related to the coupling parameter $\alpha$ in the case with $a=0$. The corresponding null geodesics (\ref{cedis}) for the polarized light becomes
\begin{eqnarray}
\frac{r^4}{(1+2\alpha)}\dot{r}^2&=&r^4E^2-(Q+L^2)\Delta,\\
r^4\dot{\theta}^2&=&Q-L^2\cot^2\theta,
\label{cedaa}
\end{eqnarray}
where $Q$ is the Carter constant. It is obvious that in this case the photon sphere for the polarized light does not depend on the coupling and is the same as that of usual Schwarzschild case without the coupling. With two impact parameters $\xi=\frac{L}{E}$ and $\sigma=\frac{Q}{E^2}$, we can obtain the two celestial coordinates
\begin{eqnarray}
\label{kztqzb}
x&=&-\lim\limits_{r_{0}\rightarrow\infty}{r}\frac{p^{\hat{\phi}}}{p^{\hat{r}}}\bigg|_{(r_{0},\theta_{0})}
=-\lim\limits_{r_{0}\rightarrow\infty}{r}\frac{\sqrt{g_{rr}}p_{\phi}}{
\sqrt{g_{\phi\phi}}p_{r}}\bigg|_{(r_{0},\theta_{0})} =-\frac{\xi}{\sin\theta_0}\nonumber\\
y&=&\lim\limits_{r_{0}\rightarrow\infty}{r}\frac{p^{\hat{\theta}}}{p^{\hat{r}}}\bigg|_{(r_{0},\theta_{0})}
=\lim\limits_{r_{0}\rightarrow\infty}{r}\frac{\sqrt{g_{rr}}p_{\theta}}{
\sqrt{g_{\theta\theta}}p_{r}}\bigg|_{(r_{0},\theta_{0})}
=\pm\sqrt{\sigma-\xi^{2}\cot^{2}\theta_{0}},
\end{eqnarray}
which are independent of the coupling constant $\alpha$. This means that the black hole shadows casted by the polarized lights are same even if their propagation paths are different in the non-rotating case. In Fig.\ref{as0}, we also plot the effect of the polarized light on the black hole shadow in the non-rotating case. Here, we adopt the operation in Refs.\cite{sw,swo,astro,chaotic,binary,sha18,my,BI,swo7,swo8,swo9,swo10} and divide the celestial sphere into four quadrants marked by different colors (green, blue, red, and yellow). The grid of longitude and latitude lines is marked with adjacent brown lines separated by $10\degree$. The distribution of these color regions and lines in can reflect the distortion of an image due to the strong gravitational lensing of black hole. The white regions in Figs.\ref{as0} are caused by lights from a reference spot lied in the line between black hole and observer, which could provide a direct demonstration of Einstein ring. The black patch is used to denote black hole shadow. Fig.\ref{as0} confirms that the black hole shadows in non-rotating case do not depends on the polarization of light and the coupling parameter $\alpha$. However, the color region distribution in Fig.\ref{as0} changes with the polarization of light and the parameter $\alpha$, which means the birefringence phenomenon occurs exactly in this case without non-vanishing $\alpha$. The radius of Einstein ring in Fig.\ref{as0}, where corresponds to the non-rotating case related to the effective metric (\ref{mec2}), can be expressed as
\begin{eqnarray}\label{EstRing}
R_E\simeq D_{OL}\theta_E=\sqrt{\frac{4M}{(1+2\alpha)^{1/2}}\frac{D_{OL}D_{LS}}{D_{OS}}},
\end{eqnarray}
where the observer-source distance $D_{OS}=D_{LS}+D_{OL}$. $D_{LS}$ and $D_{OL}$ denote the lens-source distance and the observer-lens distance, respectively. It is obvious that the radius of Einstein ring $R_E$ decreases with the increase of the coupling parameter $\alpha$, which is observed in Fig.\ref{as0}.
In a word, in the non-rotating case, the Lorentz symmetry breaking changes the propagation of polarized lights, but the black hole shadows do not contain any information on Lorentz symmetry breaking for the selected bumblebee field (\ref{beefield}).

\begin{figure}
\includegraphics[width=5.0cm ]{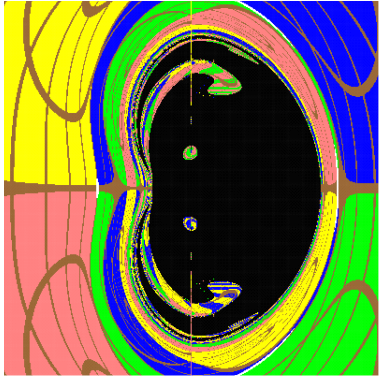}
\caption{The black hole shadow for the effective metric (\ref{mec2}) with $\alpha=-0.4$ and $a=0.998$, which shows that the main eyebrow is  disconnected from the main shadow. Here we set the black hole $M=1$ and the observer locates in the position with $r_{obs}=50M$  and  $\theta_{obs}=\pi/2$.}
\label{as6}
\end{figure}

Let us now focus on the nonzero $a$ case. It is clear that Eq.(\ref{cedis}) can not be variable-separable due to the presence of the cross term $\dot{r}\dot{\theta}$, i.e., the $r-$ component and $\theta-$ component do not decouple from each other. This  means that the dynamical system of two kind of polarized photons is non-integrable and  the shape of black hole shadow will change.  Due to the equation (\ref{cedis}) without separable variables, we must resort to "backward ray-tracing" method \cite{sw,swo,astro,chaotic,binary,sha18,my,BI,swo7,swo8,swo9,swo10} to study numerically the shadow casted by such kinds of polarized lights. In this method, the light rays are assumed to evolve from the observer backward in time and the information carried by each ray would be respectively assigned to a pixel in a final image in the observer's sky.
For the effective metric (\ref{meun}), one can expand the observer basis $\{e_{\hat{t}},e_{\hat{r}},e_{\hat{\theta}},e_{\hat{\phi}}\}$ as a form in the coordinate basis $\{ \partial_t,\partial_r,\partial_{\theta},\partial_{\phi} \}$
\begin{eqnarray}
\label{zbbh}
e_{\hat{\mu}}=e^{\nu}_{\hat{\mu}} \partial_{\nu},
\end{eqnarray}
where the transform matrix $e^{\nu}_{\hat{\mu}}$ obeys to $\gamma_{\mu\nu}e^{\mu}_{\hat{\alpha}}e^{\nu}_{\hat{\beta}}
=\eta_{\hat{\alpha}\hat{\beta}}$, and $\eta_{\hat{\alpha}\hat{\beta}}$ is the metric of Minkowski spactime. It is convenient to choose a decomposition associated with a reference frame with zero axial angular momentum in relation to spatial infinity \cite{sw,swo,astro,chaotic,binary,sha18,my,BI,swo7,swo8,swo9,swo10,wei,zero1}
\begin{eqnarray}
\label{zbbh1}
e^{\nu}_{\hat{\mu}}=\left(\begin{array}{cccc}
\zeta&0&0&\gamma\\
0&A^r&A^{r\theta}&0\\
0&0&A^{\theta}&0\\
0&0&0&A^{\phi}
\end{array}\right),
\end{eqnarray}
where $\zeta$, $\gamma$, $A^r$, $A^{\theta}$, $A^{r\theta}$,and $A^{\phi}$ are real coefficients.
From the Minkowski normalization
\begin{eqnarray}
e_{\hat{\mu}}e^{\hat{\nu}}=\delta_{\hat{\mu}}^{\hat{\nu}},
\end{eqnarray}
one can obtain
\begin{eqnarray}
\label{xs}
&&A^r=\frac{1}{\sqrt{\gamma_{rr}}},\;\;\;\;\;\;\;\;\;\;\;\;\;\;\;\;
A^{\theta}=\frac{1}{\sqrt{\gamma_{\theta\theta}}},\;\;\;\;\;\;\;\;\;\;\;\;\;\;\;
A^{\phi}=\frac{1}{\sqrt{\gamma_{\phi\phi}}},\;\;\;\;\;\;\;\;\;\;\;
\zeta=\sqrt{\frac{\gamma_{\phi \phi}}{\gamma_{t\phi}^{2}-\gamma_{tt}\gamma_{\phi \phi}}},
\nonumber\\
&&A^{r\theta}=-\frac{\gamma_{r\theta}}{\gamma_{\theta\theta}}
\sqrt{\frac{\gamma_{\theta\theta}}{\gamma_{rr}\gamma_{\theta\theta}-
\gamma_{r\theta}^{2}}},\;\;\;\;\;\;\;\;\;\;\;\;\;\;\;\;\;\;\;\; \gamma=-\frac{\gamma_{t\phi}}{\gamma_{\phi\phi}}\sqrt{\frac{\gamma_{\phi \phi}}{\gamma_{t\phi}^{2}-\gamma_{tt}\gamma_{\phi \phi}}}.
\end{eqnarray}
Therefore, one can get the locally measured four-momentum $p^{\hat{\mu}}$ of a photon by the projection of its four-momentum $p^{\mu}$  onto $e_{\hat{\mu}}$,
\begin{eqnarray}
\label{dl}
p^{\hat{t}}=-p_{\hat{t}}=-e^{\nu}_{\hat{t}} p_{\nu},\;\;\;\;\;\;\;\;\;
\;\;\;\;\;\;\;\;\;\;\;p^{\hat{i}}=p_{\hat{i}}=e^{\nu}_{\hat{i}} p_{\nu}.
\end{eqnarray}
With the help of Eq.(\ref{xs}), the locally measured four-momentum $p^{\hat{\mu}}$ can be further written as
\begin{eqnarray}
\label{kmbh}
p^{\hat{t}}&=&\zeta E-\gamma L,\;\;\;\;\;\;\;\;\;\;\;\;\;\;\;\;\;\;\;\;
p^{\hat{r}}=\frac{1}{\sqrt{\gamma_{rr}}}p_{r}-
\frac{\gamma_{r\theta}}{\gamma_{\theta\theta}}
\sqrt{\frac{\gamma_{\theta\theta}}{\gamma_{rr}\gamma_{\theta\theta}-
\gamma_{r\theta}^{2}}}p_{\theta},\nonumber\\
p^{\hat{\theta}}&=&\frac{1}{\sqrt{\gamma_{\theta\theta}}}p_{\theta},
\;\;\;\;\;\;\;\;\;\;\;\;\;\;\;\;\;\;\;\;\;\;
p^{\hat{\phi}}=\frac{1}{\sqrt{\gamma_{\phi\phi}}}L.
\end{eqnarray}
Repeating the similar operations in Refs.\cite{sw,swo,astro,chaotic,binary,sha18,my,BI,swo7,swo8,swo9,swo10},
one can obtain the position of photon's image in observer's sky
\begin{eqnarray}
\label{xd1}
x&=&-r_{obs}\frac{p^{\hat{\phi}}}{p^{\hat{r}}}
=-r_{obs}\frac{\sqrt{(F^2_i-H^2_i)F_i\Delta}L}{
[\sqrt{F^2_i-H^2_i}(F_i\dot{r}-H_i\sqrt{\Delta}\dot{\theta})+
H_i(F_i\sqrt{\Delta}\dot{\theta}-H_i\dot{r})]\Sigma\sin\theta}, \nonumber\\
y&=&r_{obs}\frac{p^{\hat{\theta}}}{p^{\hat{r}}}=
r_{obs}\frac{\sqrt{F^2_i-H^2_i}(F_i\sqrt{\Delta}\dot{\theta}-H_i\dot{r})}
{\sqrt{F^2_i-H^2_i}(F_i\dot{r}-H_i\sqrt{\Delta}\dot{\theta})+
H_i(F_i\sqrt{\Delta}\dot{\theta}-H_i\dot{r})},
\end{eqnarray}
where the spatial position of observer is set to ($r_{obs}, \theta_{obs}$).
\begin{figure}
  \includegraphics[width=16.0cm ]{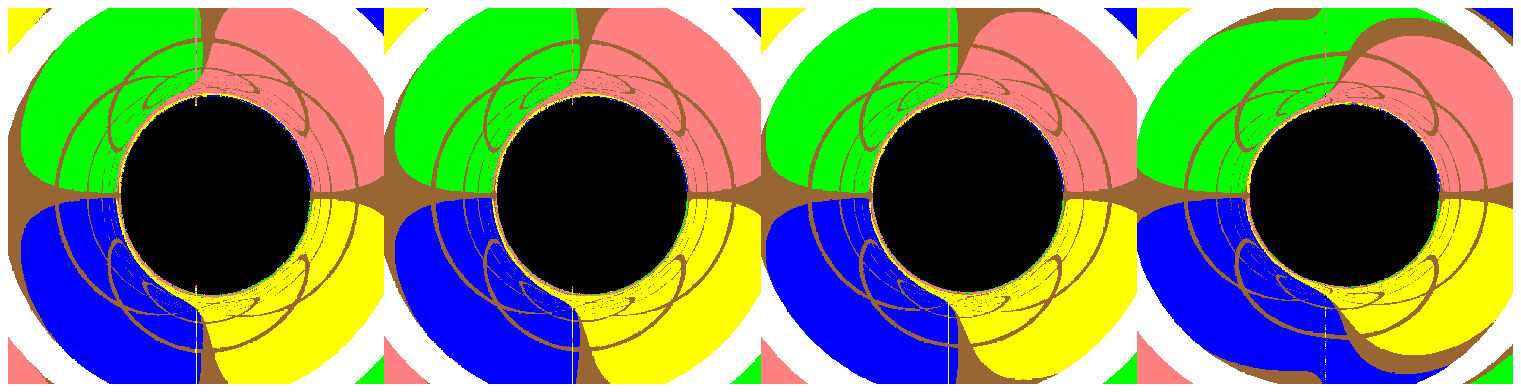}
\caption{The change of black hole shadow with the coupling constant $\alpha$ for the metric (\ref{mec1}). Here we set the black hole $M=1$ and rotation parameter $a=0.5$, and the observer locates in the position with $r_{obs}=50M$  and  $\theta_{obs}=\pi/2$. The figures from left to right correspond to $\alpha=-0.3$, $-0.2$, $0$, and $0.2$, respectively.}
\label{as1}
\end{figure}
\begin{figure}
 \includegraphics[width=16.0cm ]{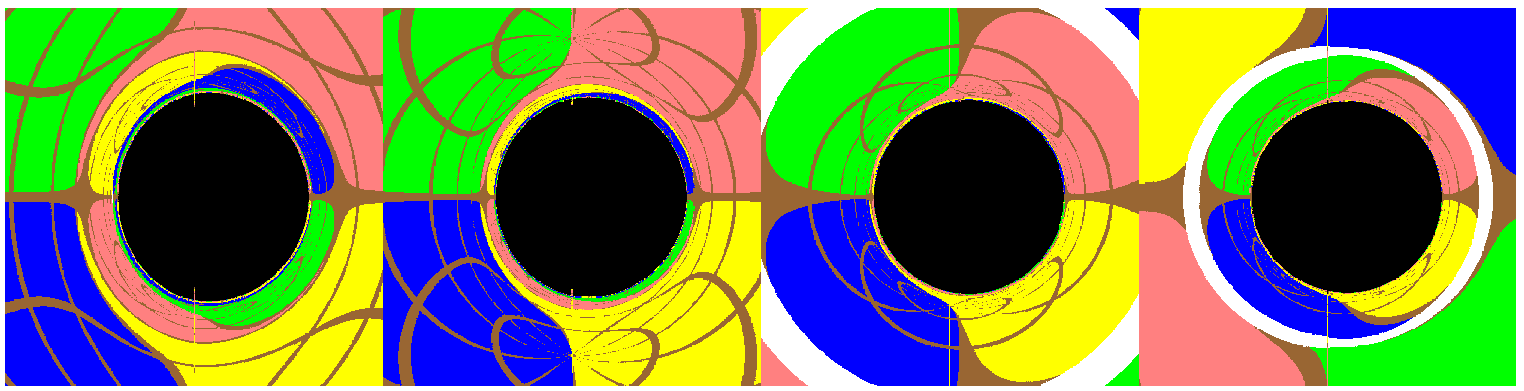}
\caption{The change of black hole shadow with the coupling constant $\alpha$ for the metric (\ref{mec2}). Here we set the black hole $M=1$ and rotation parameter $a=0.5$, and the observer locates in the position with $r_{obs}=50M$  and  $\theta_{obs}=\pi/2$. The figures from left to right correspond to $\alpha=-0.3$, $-0.2$, $0$, and $0.2$, respectively.}
\label{as2}
\end{figure}
\begin{figure}
 \includegraphics[width=16.0cm ]{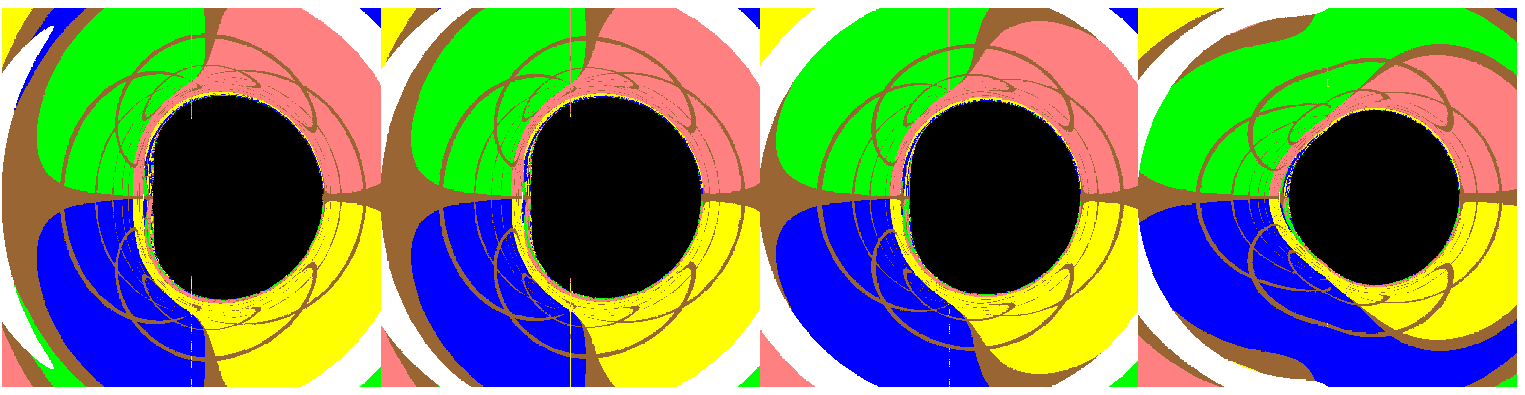}
\caption{The change of black hole shadow with the coupling constant $\alpha$ for the metric (\ref{mec1}). Here we set the black hole $M=1$ and rotation parameter $a=0.998$, and the observer locates in the position with $r_{obs}=50M$  and  $\theta_{obs}=\pi/2$. The figures from left to right correspond to $\alpha=-0.3$, $-0.2$, $0$, and $0.2$, respectively.}
\label{as3}
\end{figure}
\begin{figure}
\includegraphics[width=16.0cm ]{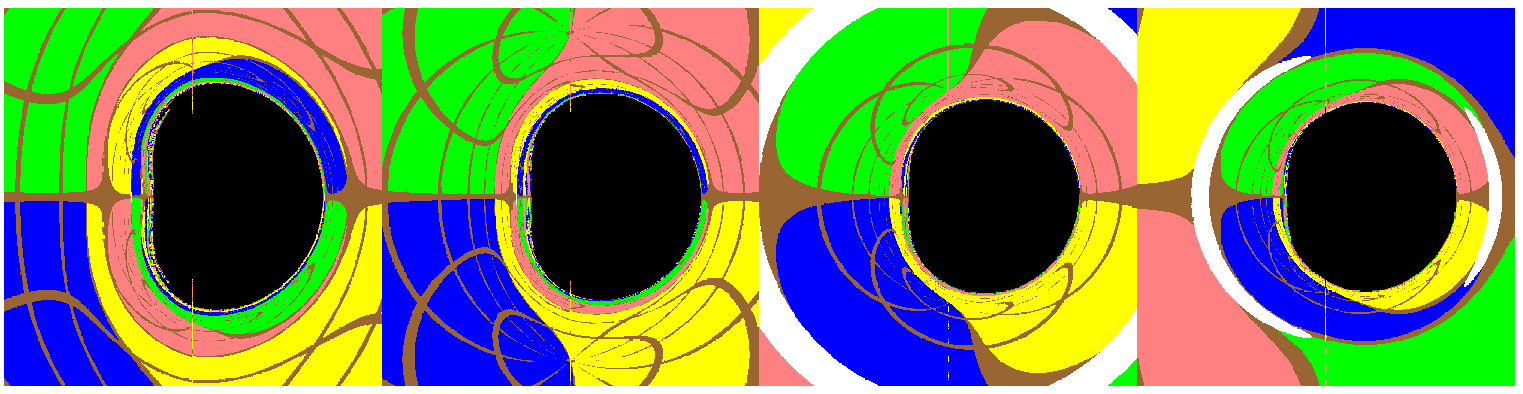}
\caption{The change of black hole shadow with the coupling constant $\alpha$ for the metric (\ref{mec2}). Here we set the black hole $M=1$ and rotation parameter $a=0.998$, and the observer locates in the position with $r_{obs}=50M$  and  $\theta_{obs}=\pi/2$. The figures from left to right correspond to $\alpha=-0.3$, $-0.2$, $0$, and $0.2$, respectively.}
\label{as4}
\end{figure}
\begin{figure}
\includegraphics[width=5.0cm ]{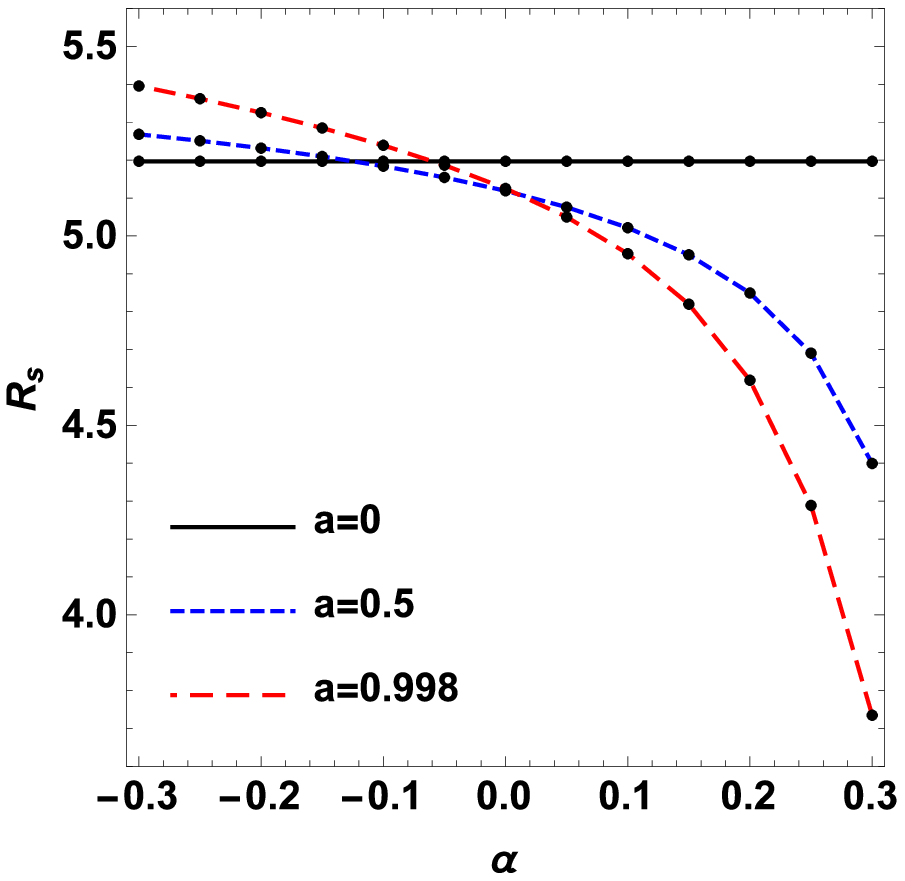}\quad\quad\includegraphics[width=5.0cm ]{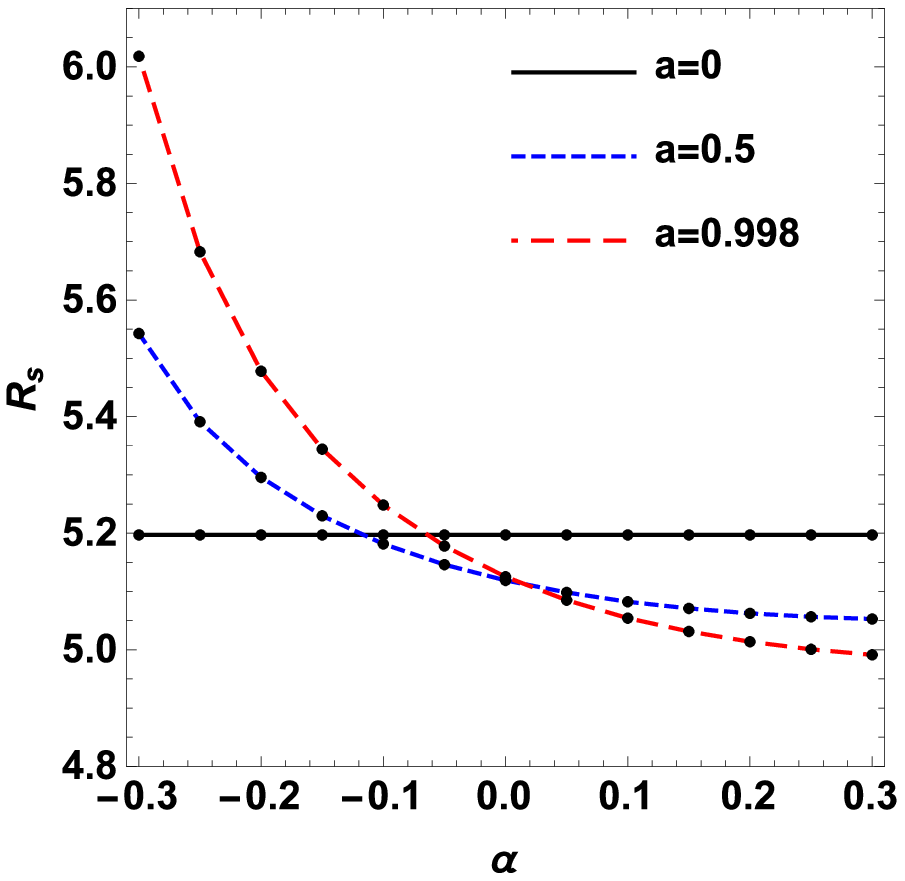}\\
\includegraphics[width=5.0cm ]{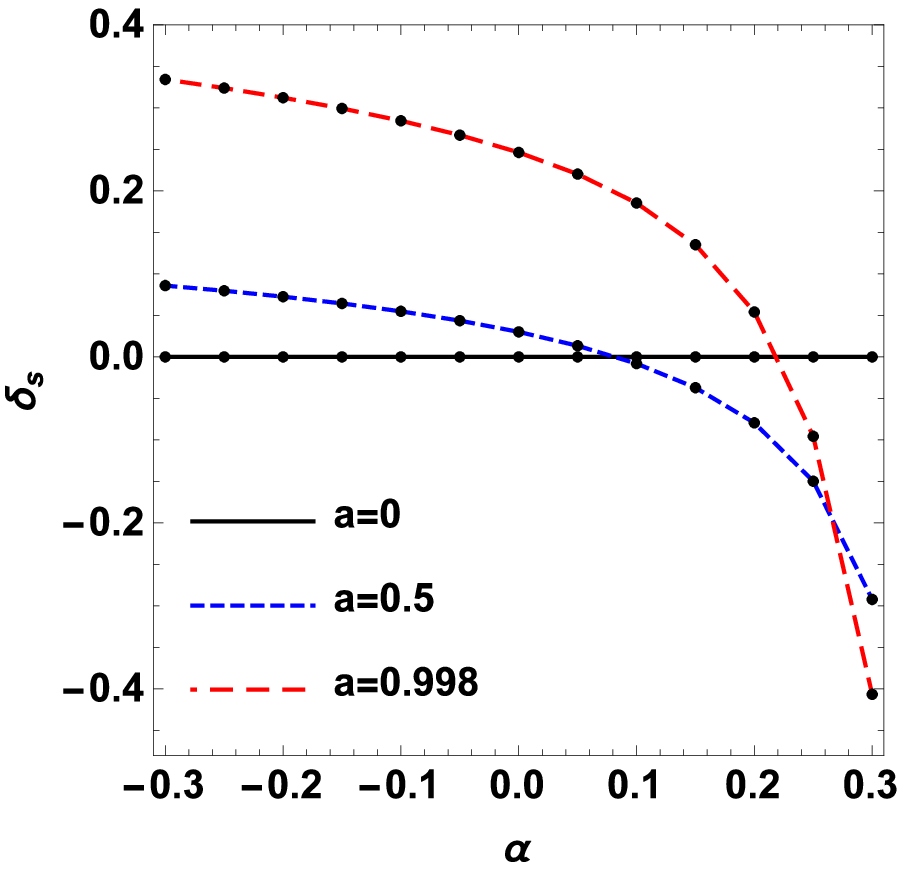}\quad\quad\includegraphics[width=5.0cm ]{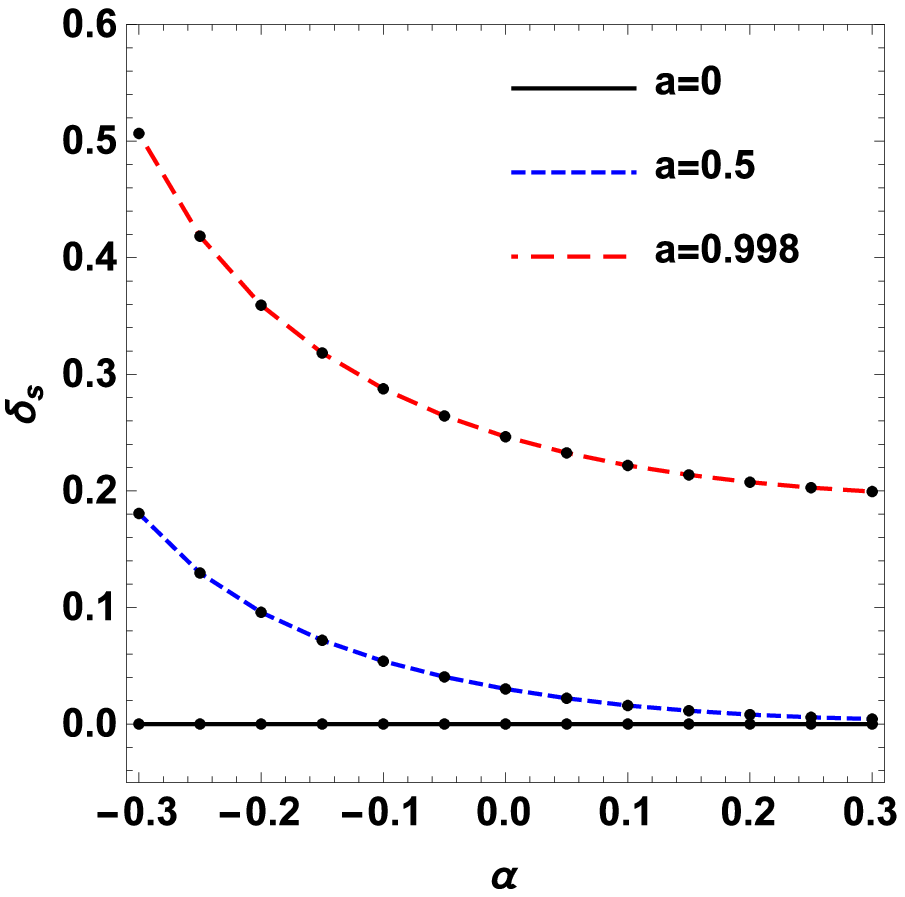}
\caption{The change of the black hole shadow size $R_s$ and the distortion parameter $\delta_s$ with the coupling parameter $\alpha$ for the different rotation parameter $a$. The left and right panels correspond to the effective metrics (\ref{mec1}) and (\ref{mec2}), respectively. }
\label{as40}
\end{figure}
\begin{figure}
\includegraphics[width=16.0cm ]{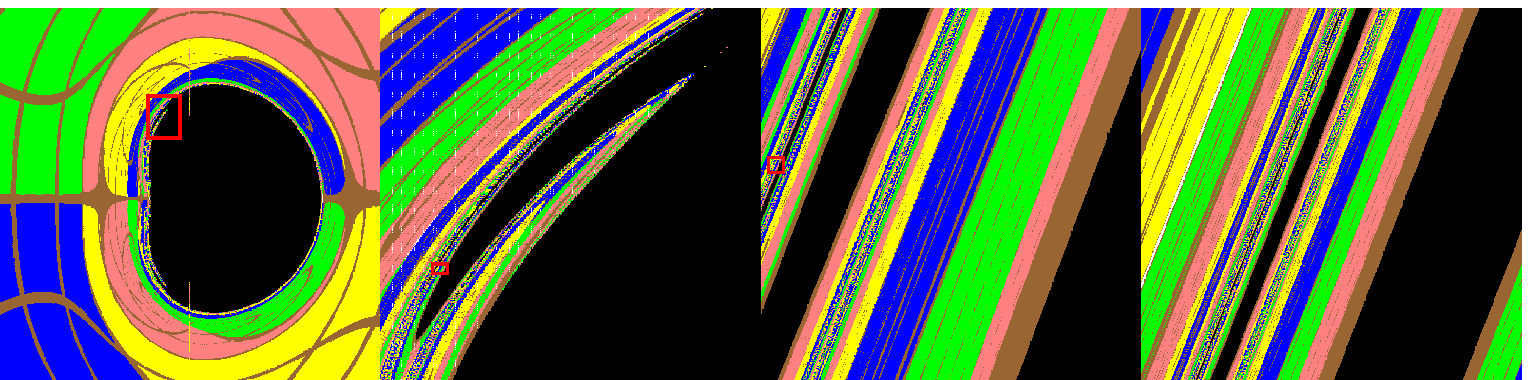}
\caption{The enlargement of the left panel in Fig.\ref{as4}. It shows that there exist the eyebrow shape shadow and the self-similar fractal structures in black hole shadow.}
\label{as5}
\end{figure}
In Figs.\ref{as1}-\ref{as4}, we present  black hole shadow for the different coupling constant $\alpha$ and rotation parameter $a$. Comparing the color regions distributions with the same parameters in  Figs.\ref{as1}-\ref{as2} or Figs.\ref{as3}-\ref{as4}, we find that the propagation of light in the black hole spacetime also depends on the polarization of light, which is similar to those in the non-rotating case. However, in the rotating case, the black hole shadow also depends on the coupling constant  $\alpha$ and the polarization of light, which differs from those in the previous non-rotating case. From Figs.\ref{as1}-\ref{as4}, we find that the negative coupling constant $\alpha$ elongates the shadow along $y-$axial direction for the fixed rotation parameter $a$ in the both cases (\ref{mec1}) and (\ref{mec2}), but the positive $\alpha$ squeezes the shadow along $y-$axis. In Fig.\ref{as40}, we also plot the change of the black hole shadow size parameter $R_s$ and the distortion parameter $\delta_s$ with the coupling constant $\alpha$. Here, $R_s$ and $\delta_s$, firstly defined in Ref. \cite{kk}, can be expressed as
\begin{eqnarray}
R_s=\frac{(x_t-x_r)^2+y^2_t}{2(x_r-x_t)},\quad\quad\quad\quad\delta_s=2-\frac{x_r-x_l}{R_s},
\end{eqnarray}
where the point $(x_t,y_t)$ is located at the top position of the shadow. The points
$(x_r, 0)$ and $(x_l,0)$ lie, respectively, at the most right position and the most left position of shadow along the horizontal line $y=0$. From Fig.\ref{as40}, we find that
the negative coupling constant $\alpha$ make the shadow size increase
and positive one make the shadow size decrease for fixed rotation parameter $a$.
Moreover,  we find that the shadow casted by the polarized light propagating in the effective metric (\ref{mec1}) has a smaller distortion parameter $\delta_s$ than in the effective metric (\ref{mec2}). With the increase of the rotation parameter, the effect of polarized lights on the black hole shadow becomes more distinctly. It is obvious that the effect of Lorentz symmetry breaking on black hole shadow increases with the absolute value $\alpha$ in this case.
Moreover, from the left panel in Fig.\ref{as4}, we can find  the eyebrow shape shadow and the self-similar fractal structures of black hole shadow arising from the chaotic lensing of polarized lights, which is also shown in Fig.\ref{as5}. We also note that the main eyebrow is connected to the main shadow in Fig.\ref{as4} with $\alpha=-0.3$, but it is disconnected from the main shadow by a bright vertical line in Fig.\ref{as6} with $\alpha=-0.4$.
These features of black hole shadow casted by polarized lights could help us to understand deeply the bumblebee vector fields with Lorentz symmetry breaking and their corresponding interactions with electromagnetic field.

\section{summary}

In this paper, we have investigated firstly the equation of motion for the
photon coupled to the three special kinds of bumblebee vector fields in a Kerr black hole spacetime and then probed further the black hole shadow. In the non-rotating case, we find that the black hole shadow is independent of the polarization of light and the coupling constant even if the polarized lights propagate along  different paths. This means that Lorentz symmetry breaking arising from bumblebee vector field (\ref{beefield}) does not affect the black hole shadow in these two special cases. However, in the rotating case, the status is changed because the black hole shadow also depends on the polarization of light and the coupling constant $\alpha$. The negative coupling constant $\alpha$ elongates the shadow along $y-$ axial direction, and make the shadow size increase for the fixed rotation parameter $a$. The positive $\alpha$ squeezes the shadow along $y-$axis, but the shadow size is decreased.  Moreover,  we find that the shadow casted by the polarized light propagating in the effective metric (\ref{mec1}) has a smaller distortion parameter $\delta_s$ than in the effective metric (\ref{mec2}).  The effect of polarized lights on the black hole shadow becomes more distinctly in the case with the quicker rotation and the stronger coupling. With the development of the new generation Event Horizon Telescope, these features of black hole shadow casted by polarized lights may be detected in the future and could help us to understand the bumblebee vector field with Lorentz symmetry breaking and its interaction with electromagnetic field in the nature.

\section{\bf Acknowledgments}

This work was  partially supported by the National Natural Science
Foundation of China under Grant No.11875026, the Scientific Research Fund of Hunan Provincial Education Department Grant No. 17A124; J. Jing's work was
partially supported by the National Natural Science Foundation of
China under Grant No.11875025.


\vspace*{0.2cm}
\begin{thebibliography}{99}
\baselineskip=0.5 cm
\bibitem{fbhs1} The Event Horizon Telescope Collaboration, \textit{First M87 Event Horizon Telescope Results. I. The Shadow of the Supermassive Black Hole}, Astrophys. J. Lett. {\bf875}, L1 (2019).
\bibitem{fbhs6} The Event Horizon Telescope Collaboration, \textit{First M87 Event Horizon Telescope Results. VI. The Shadow and Mass of the Central Black Hole}, Astrophys. J. Lett. {\bf875}, L6 (2019).

 \bibitem{extr1} S. Vagnozzi and L. Visinelli, \textit{Hunting for extra dimensions in the shadow of M87*}, Phys. Rev. D {\bf100}, 024020 (2019).
\bibitem{extr2} I. Banerjee, S. Chakraborty, and S. SenGupta, \textit{Silhouette of M87*: A new window to peek into the world of hidden dimensions}, Phys. Rev. D {\bf 101}, 041301(R) (2020).
\bibitem{tomoch} Y. Chen, J. Shu, X. Xue, Q. Yuan, Y. Zhao, \textit{Probing Axions with Event Horizon Telescope Polarimetric Measurements}, Phys. Rev. Lett. {\bf124}, 061102 (2020).
\bibitem{dark1} R. Konoplya, \textit{Shadow of a black hole surrounded by dark matter}, Phys. Lett. B {\bf 795}, 1 (2019).
\bibitem{dark2}  X. Hou, Z. Xu, M. Zhou, J. Wang, \textit{Black Hole Shadow of Sgr A*   in Dark Matter Halo}, J. Cosmol. Astropart. Phys. {\bf1807}, 015, (2018).
\bibitem{dark3}  K. Jusufi, M. Jamil, P. Salucci, T. Zhu, S. Haroon, \textit{Black Hole Surrounded by a Dark Matter Halo in the M87 Galactic Center and its Identification with Shadow Images},  Phys. Rev. D {\bf100}, 044012 (2019).
\bibitem{dark4} P. V. Cunha, C. A. Herdeiro and E. Radu, \textit{EHT constraint on the ultralight scalar hair of the M87 supermassiveblack hole}, Universe {\bf5}, 220 (2019), [arXiv:1909.08039 [gr-qc]

\bibitem{epb}C. Li, S.Yan, L.Xue, X.Ren, Y.Cai, D.A. Easson, Y. Yuan, and H. Zhao, \textit{Testing the equivalence principle via the shadow of black holes}, arXiv:1912.12629 [astro-ph].

\bibitem {sb1} Y. Huang, S. Chen, and J. Jing, \textit{Double shadow of a regular phantom black hole as photons couple to Weyl tensor}, Eur. Phys. J. C {\bf76}, 594 (2016).

\bibitem{lvia1}G. Zatsepin, V. Kuzmin, \textit{Upper limit of the spectrum of cosmic rays }, JETP Lett. {\bf4}, 78 (1966).
\bibitem{lvia2} M. Takeda, \textit{et al}, \textit{Extension of the Cosmic-Ray Energy Spectrum beyond the Predicted Greisen-Zatsepin-Kuzmin Cutoff}, Phys. Rev. Lett. {\bf81}, 1163 (1998).
\bibitem{casa} R. Casana, A. Cavalcante, F. P. Poulis, and E. B. Santos, \textit{Exact Schwarzschild-like solution in a bumblebee gravity model}, Phys.
Rev. D {\bf97}, 104001 (2018).

\bibitem{kost} V. Kostelecky and S. Samuel,\textit{Gravitational phenomenology in higher-dimensional theories and strings}, Phys. Rev. D {\bf40}, 1886 (1989).
\bibitem{ber} O. Bertolami and J. Paramos, \textit{Vacuum solutions of a gravity model with vector-induced spontaneous Lorentz symmetry breaking}, Phys. Rev. D {\bf72}, 044001 (2005).
\bibitem{kost2} Q. Bailey and V. A. Kostelecky, \textit{Signals for Lorentz violation in post-Newtonian gravity}, Phys. Rev. D {\bf74}, 045001 (2006).
\bibitem{blum} R. Bluhm, N. Gagne, R. Potting, and A. Vrublevskis, \textit{Constraints and stability in vector theories with spontaneous Lorentz violation}, Phys.
Rev. D {\bf77}, 125007 (2008).
\bibitem{kost3} V. Kostelecky and J. D. Tasson, \textit{Prospects for Large Relativity Violations in Matter-Gravity Couplings}, Phys. Rev. Lett. {\bf102}, 010402
(2009).
\bibitem{seif} M. Seifert, \textit{Generalized bumblebee models and Lorentz-violating electrodynamics}, Phys. Rev. D {\bf81}, 065010 (2010).
\bibitem{malu} R.  Maluf, C. Almeida, R. Casana, and M.  Ferreira,\textit{Einstein-Hilbert graviton modes modified by the Lorentz-violating bumblebee field},
 Phys. Rev. D {\bf90}, 025007 (2014).
\bibitem{guio} G. Guiomar and J. Paramos, \textit{Astrophysical constraints on the bumblebee model}, Phys. Rev. D {\bf90}, 082002 (2014).
\bibitem{esco}  C. Escobar and A. MartinRuiz, \textit{Equivalence between bumblebee models and electrodynamics in a nonlinear gauge}, Phys. Rev. D {\bf95}, 095006
(2017).
\bibitem{assu} J. Assuncao, T. Mariz, J. Nascimento, and A. Petrov, \textit{Dynamical Lorentz symmetry breaking in a tensor bumblebee model}, Phys. Rev. D {\bf100}, 085009 (2019).


\bibitem{ovgu} A. Ovgun, K. Jusufi, and I. Sakalli, \textit{Gravitational Lensing Under the Effect of Weyl and Bumblebee Gravities: Applications of Gauss-Bonnet Theorem}, Ann. Phys. (Amsterdam) {\bf399}, 193 (2018).
\bibitem{kanz} S. Kanzi and I. Sakalli, \textit{GUP Modified Hawking Radiation in Bumblebee Gravity}, Nucl. Phys. B {\bf946}, 114703 (2019).

\bibitem{ding} C. Ding, C. Liu, R. Casana, and A. Cavalcante, \textit{Exact Kerr-like solution and its shadow in a gravity model with spontaneous Lorentz symmetry breaking}, Eur. Phys. J. C {\bf 80}, 178 (2020).
\bibitem{ding1}  C. Liu, C. Ding, and J. Jing, \textit{Thin accretion disk around a rotating Kerr-like black hole in Einstein-bumblebee gravity model }, arXiv:1910.13259.
\bibitem{liz} Z. Li, A. Ovgun, \textit{Finite-distance gravitational deflection of massive particles by a Kerr-like black hole in the bumblebee gravity model},Phys. Rev. D {\bf101}, 024040 (2020).
\bibitem{ovgu2}A. Ovgun, K. Jusufi, and I. Sakalli, \textit{Exact traversable wormhole solution in bumblebee gravity}, Phys. Rev. D {\bf99}, 024042 (2019).

\bibitem{cape}D. Capelo, J. Paramos, \textit{Cosmological implications of bumblebee vector models}, Phys. Rev. D {\bf91}, 104007 (2015).



\bibitem{tomo} T. Fujita, R. Tazaki, K. Toma, \textit{Hunting Axion Dark Matter with Protoplanetary Disk Polarimetry},Phys. Rev. Lett. {\bf122}, 191101 (2019).

\bibitem{Alexis} A. Plascencia,  A. Urbano, \textit{Black hole superradiance and polarization-dependent bending of light},
J. Cosmol. Astropart. Phys. {\bf04} 059 (2018), arxiv: 1711.08298 [gr-qc].


\bibitem{Drummond}I. T. Drummond and S. J. Hathrell, \textit{QED vacuum polarization in a background gravitational field and its effect on the velocity of photons}, Phys. Rev. D {\bf22}, 343
(1980).
\bibitem{Daniels} R. D. Daniels, and G. M. Shore,  \textit{``Faster than light" photons and charged black holes }, Nucl. Phys. B {\bf425}, 634 (1994).

\bibitem{Daniels1}  R. D. Daniels, and G. M. Shore, \textit{``Faster than light" photons and rotating black holes }, Phys. Lett. B {\bf367}, 75 (1996).

\bibitem{Caip} R. G. Cai, \textit{Propagation of vacuum polarized photons in topological black hole spacetimes }, Nucl. Phys. B {\bf 524}, 639 (1998).
\bibitem{Cho1}H. T. Cho,  \textit{``Faster than light" photons in dilaton black hole spacetimes}, Phys. Rev. D {\bf56}, 6416 (1997).

\bibitem{Lorenci} V. A. De Lorenci, R. Klippert, M. Novello, and J. M. Salim, \textit{Light propagation in non linear electrodynamics},Phys.Lett. B {\bf482}, 134 (2000).

\bibitem{Lorenci1}D. A. R. Dalvit, F. D. Mazzitelli, and C. Molina-Paris, \textit{One-loop graviton corrections to Maxwell¡¯s equations}, Phys. Rev. D {\bf 63}, 084023 (2001).

 \bibitem{Lorenci2} N. Ahmadi and M. N. Zonoz, \textit{Quantum gravitational optics: the induced phase}, Class. Quant. Grav. {\bf25}, 135008 (2008).
\bibitem{Breton} N. Breton, \textit{Born-Infeld generalization of the Reissner-Nordstrom black hole}, Class. Quantum Grav. {\bf19}, 601 (2002).

\bibitem{sw} P. V. P. Cunha, C. Herdeiro, E. Radu and H. F. Runarsson, \textit{Shadows of Kerr black holes with scalar hair}, Phys. Rev. Lett. {\bf115}, 211102 (2015), arXiv:1509.00021;
\bibitem{swo} P. V. P. Cunha, C. Herdeiro, E. Radu and H. F. Runarsson, \textit{Shadows of Kerr black holes with and without scalar hair}, Int. J. Mod. Phys. D {\bf25}, 1641021 (2016), arXiv:1605.08293.
\bibitem{astro}F. H. Vincent, E. Gourgoulhon, C. Herdeiro and E. Radu, \textit{Astrophysical imaging of Kerr black holes with scalar hair}, Phys. Rev. D {\bf94}, 084045 (2016), arXiv:1606.04246.
\bibitem{chaotic} P. V. P. Cunha, J. Grover, C. Herdeiro, E. Radu, H. Runarsson, and A. Wittig, \textit{Chaotic lensing around boson stars and Kerr black holes with scalar hair}, Phys. Rev. D {\bf94}, 104023 (2016).
\bibitem{binary} J. O. Shipley, and S. R. Dolan, \textit{Binary black hole shadows, chaotic scattering and the Cantor set}, Class. Quantum Grav. {\bf33}, 175001 (2016).
\bibitem{sha18} A. Bohn, W. Throwe, F. Hbert, K. Henriksson, and D. Bunandar, \textit{What does a binary black hole merger look like?}, Class. Quantum Grav. {\bf32}, 065002 (2015), arXiv: 1410.7775.
\bibitem{my} M. Wang, S. Chen, J. Jing, \textit{Shadows of a compact object with magnetic dipole by chaotic lensing}, Phys. Rev. D {\bf97}, 064029 (2018).
 \bibitem{BI} J. Grover, A. Wittig, \textit{Black Hole Shadows and Invariant Phase Space Structures}, Phys. Rev. D {\bf96}, 024045 (2017).
\bibitem{swo7} T. Johannsen, \textit{Photon Rings around Kerr and Kerr-like Black Holes}, Astrophys. J. {\bf777}, 170, (2013).
\bibitem{swo8} R. Roy, U. Yajnik, \textit{Evolution of black hole shadow in the presence of ultralight bosons}, Phys. Lett. B {\bf 803}, 135284 (2020).
\bibitem{swo9} Z. Younsi, A. Zhidenko, L.Rezzolla, R. Konoplya and Y. Mizuno, \textit{New method for shadow calculations: Application to parametrized axisymmetric black holes}, Phys. Rev. D {\bf 94}, 084025 (2016).
 \bibitem{swo10} M. Wang, S. Chen and J. Jing, \textit{ Chaotic shadow of a non-Kerr rotating compact object with quadrupole mass}, Phys. Rev. D {\bf 98} 104040,  (2018).
 \bibitem{wei} S. Wei, Y. Liu, \textit{Testing the nature of Gauss-Bonnet gravity by four-dimensional rotating black hole shadow}, arXiv:2003.07769.

\bibitem{zero1} V. Frolov and I. Novikov, \textit{Black Hole Physics: Basic concepts and new developments}, (Kluwer Academic Publishers, 1998).
\bibitem{kk} K. Hioki and K. I. Maeda, \textit{Measurement of the Kerr Spin Parameter by Observation of a Compact Object¡¯s Shadow}, Phys. Rev. D {\bf80}, 024042 (2009), arXiv:0904.3575.



\end{thebibliography}
\end{document}